\documentclass{elsart}
\usepackage{graphics}
\usepackage{epsfig}
\usepackage{amssymb}
\begin{document}
\newcommand{\EP}{\mbox{e$^+$}}
\newcommand{\EM}{\mbox{e$^-$}}
\newcommand{\EPEM}{\mbox{e$^+$e$^-$}}
\newcommand{\EMEM}{\mbox{e$^-$e$^-$}}
\newcommand{\EE}{\mbox{ee}}
\newcommand{\GG}{\mbox{$\gamma\gamma$}}
\newcommand{\GP}{\mbox{$\gamma$e$^+$}}
\newcommand{\GE}{\mbox{$\gamma$e}}
\newcommand{\LGE}{\mbox{$L_{\GE}$}}
\newcommand{\LGG}{\mbox{$L_{\GG}$}}
\newcommand{\LEE}{\mbox{$L_{\EE}$}}
\newcommand{\TEV}{\mbox{TeV}}
\newcommand{\WGG}{\mbox{$W_{\gamma\gamma}$}}
\newcommand{\GEV}{\mbox{GeV}}
\newcommand{\EV}{\mbox{eV}}
\newcommand{\CM}{\mbox{cm}}
\newcommand{\M}{\mbox{m}}
\newcommand{\MM}{\mbox{mm}}
\newcommand{\NM}{\mbox{nm}}
\newcommand{\MKM}{\mbox{$\mu$m}}
\newcommand{\E}{\mbox{$\epsilon$}}
\newcommand{\EN}{\mbox{$\epsilon_n$}}
\newcommand{\EI}{\mbox{$\epsilon_i$}}
\newcommand{\ENI}{\mbox{$\epsilon_{ni}$}}
\newcommand{\ENX}{\mbox{$\epsilon_{nx}$}}
\newcommand{\ENY}{\mbox{$\epsilon_{ny}$}}
\newcommand{\EX}{\mbox{$\epsilon_x$}}
\newcommand{\EY}{\mbox{$\epsilon_y$}}
\newcommand{\SEC}{\mbox{s}}
\newcommand{\CMS}{\mbox{cm$^{-2}$s$^{-1}$}}
\newcommand{\MRAD}{\mbox{mrad}}
\newcommand{\IND}{\hspace*{\parindent}}
\newcommand{\beq}{\begin{equation}}
\newcommand{\eeq}{\end{equation}}
\newcommand{\beqn}{\begin{eqnarray}}
\newcommand{\eeqn}{\end{eqnarray}}
\newcommand{\dst}{\displaystyle}
\newcommand{\bm}{\boldmath}
\newcommand{\BX}{\mbox{$\beta_x$}}
\newcommand{\BY}{\mbox{$\beta_y$}}
\newcommand{\BI}{\mbox{$\beta_i$}}
\newcommand{\SX}{\mbox{$\sigma_x$}}
\newcommand{\SY}{\mbox{$\sigma_y$}}
\newcommand{\SZ}{\mbox{$\sigma_z$}}
\newcommand{\SI}{\mbox{$\sigma_i$}}
\newcommand{\SIP}{\mbox{$\sigma_i^{\prime}$}}
\newcommand{\n}{\mbox{$n_f$}}
\begin{frontmatter}
\title{Photon collider at TESLA \thanksref{title}}  
\author{Valery Telnov \thanksref{auth}}
\address{Institute of Nuclear Physics, 630090 Novosibirsk, Russia \\
and DESY, Notkestr.85, D-22603 Hamburg, Germany} \date{}
\thanks[title]{Talk at the Inter. Workshop on
High Energy Photon Colliders, Hamburg, June 14--17, 2000}
\thanks[auth]{e-mail:telnov@inp.nsk.su, check via e:mail current address.}
\begin{abstract}
 
  High energy photon colliders (\GG, \GE) based on backward Compton
  scattering of laser light is a very natural addition to \EPEM\
  linear colliders. In this report we consider this option for the
  TESLA project.  Recent study has shown that the horizontal emittance
  in the TESLA damping ring can be further decreased by a factor of
  four.  In this case the \GG\ luminosity in the high energy part of
  spectrum can reach about $(1/3)L_{\EPEM}$.  Typical cross sections
  of interesting processes in \GG\ collisions are higher than those in
  \EPEM collisions by about one order of magnitude, so the number of
  events in \GG\ collisions will be more than that in \EPEM\
  collisions.  Photon colliders can, certainly, give additional
  information and they are the best for the study of many phenomena.
  The main question is now the technical feasibility. The key new
  element in photon colliders is a very powerful laser system.  An
  external optical cavity is a promising approach for the TESLA
  project. A free electron laser is another option.  However, a more
  straightforward solution is ``an optical storage ring (optical
  trap)'' with a diode pumped solid state laser injector which is
  today technically feasible.  This paper briefly reviews the status
  of a photon collider based on the linear collider TESLA, its
  possible parameters and existing problems.

\vspace*{0cm}
\end{abstract}
\end{frontmatter}

\vspace{-1cm}
\section{Introduction}

Over the last decade, several laboratories in the world have been
working on linear \EPEM\ collider projects with an energy from several
hundreds GeV up to several TeV: these are NLC(USA)~\cite{NLC},
JLC(Japan)~\cite{JLC}, TESLA(Europe)~\cite{TESLA}, CLIC
(CERN)~\cite{CLIC}. In addition to the \EPEM\ physics program, linear
colliders provide a unique opportunity to study \GG\ and \GE\
interactions at energies and luminosities comparable to those in
\EPEM\ collisions~\cite{GKST81,GKST83,GKST84,TEL90}.  High
energy photons for \GG, \GE\ collisions can be obtained using Compton
backscattering of laser light off the high energy electrons.

\begin{figure}[!hbt]
\centering
\vspace*{0.7cm}
\epsfig{file=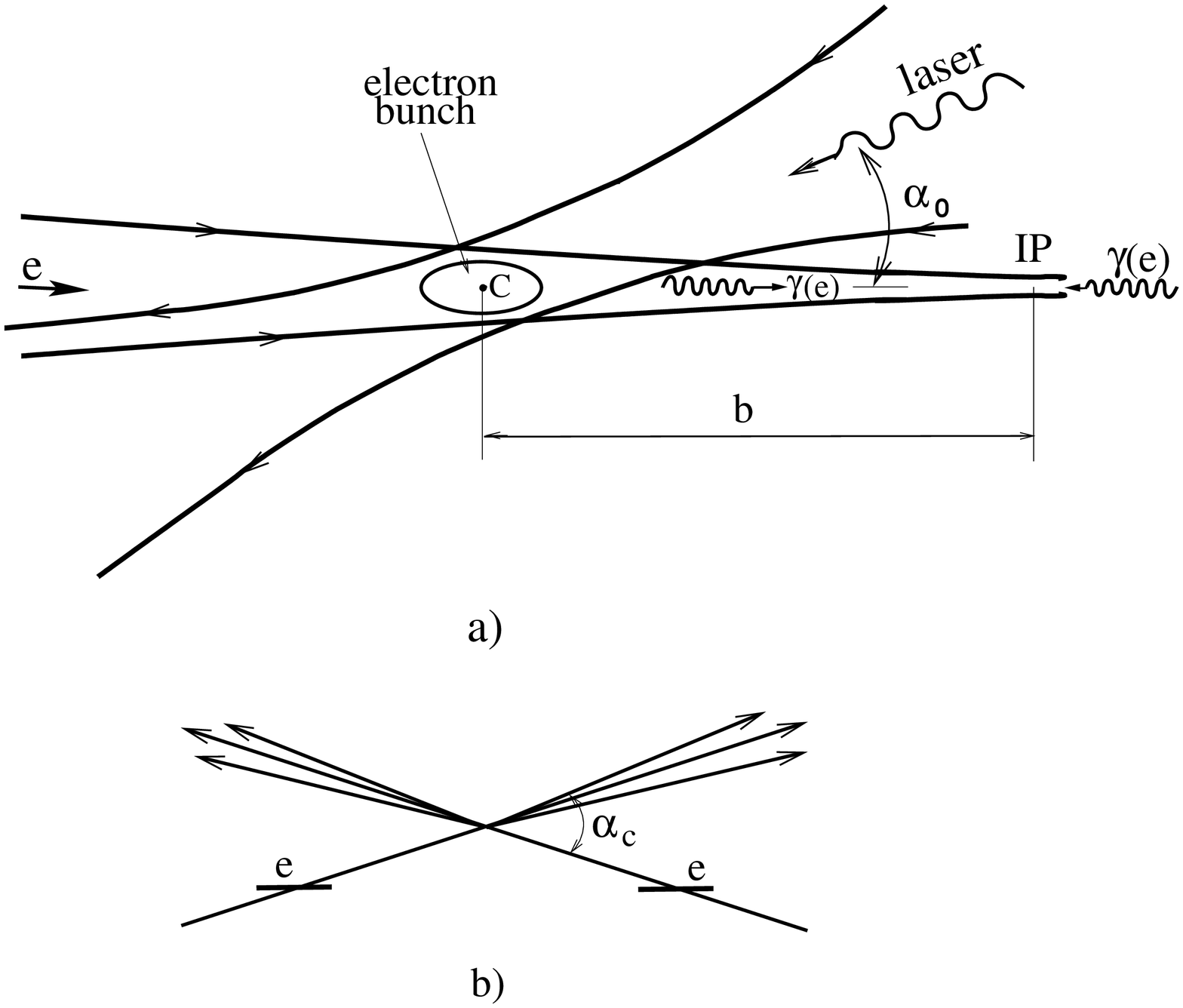,width=10cm,angle=0}
\vspace*{-0.cm}
\caption{Scheme of  \GG, \GE\ collider.}
\label{ggcol}
\end{figure}

The basic scheme of a photon collider is shown in
Fig.~\ref{ggcol}. Two electron beam of energy $E_0$ after the final
focus system travel towards the interaction point (IP) and at a
distance $b$ of about 0.1--0.5 cm from the IP collide with the focused
laser beam.  After scattering, the photons have an energy close to
that of the initial electrons and follow their direction to the
interaction point (IP) (with some small additional angular spread of
the order of $1/\gamma$, where $\gamma = E_0/mc^2$), where they
collide with a similar opposite high energy photons or electrons.
Using a laser flash energy of several Joules one can ``convert''
almost all electrons to high energy photons. The photon spot size at
the IP will be almost equal to that of the electron at the IP would
have and therefore the luminosity of \GG, \GE\ collisions will be of
the same order as the ``geometric'' luminosity of the basic \EMEM\
beams. To avoid background from the disrupted beams, 
a crab crossing scheme is used (Fig.1b).

 The maximum energy of the scattered photons is
\begin{equation}
\omega_m=\frac{x}{x+1}E_0; \;\;\;\;
x = \frac{4E_0\omega_0\cos^2{\alpha_0/2}}{m^2c^4}
 \simeq 15.3\left[\frac{E_0}{\TEV}\right]
\left[\frac{\omega_0}{eV}\right],
\end{equation}
where $E_0$ is the electron beam energy, $\omega_0$ the energy of the laser
photon, $\alpha_0$ is the angle between electron and laser beam (see,
Fig.\ref{ggcol}a).  
For example: $E_0$ =250\,\, GeV, $\omega_0 =1.17$ eV ($\lambda=1.06$
\MKM, Nd:glass and other powerful solid state lasers) we obtain
$x=4.5$ and $\omega/E_0 = 0.82$.

For increasing values of $x$ the high energy photon spectrum becomes
more peaked.  However, at $x > 4.8$ the high energy photons are lost
due to \EPEM\ creation in the collisions with laser
photons~\cite{GKST83,TEL90,TEL95}. The optimum laser wave length
$\lambda[\MKM\ ] \sim 4E_0[\TEV\ ].$ So, for \GG\ collisions the maximum
energy is $z_m = W_{\GG,max}/2E_0 = x/(x+1)\sim 0.8$ and in \GE\
collisions $z_m = W_{\GE,max}/2E_0= \sqrt{x/x+1)} \sim 0.9$.  For an
introduction to photon colliders see
refs~\cite{GKST83,GKST84,TEL90,TEL95,BERK,Brinkmann98}, and
introductory talks at this Workshop~\cite{TELgoals,Takahashi}.

Below  most important issues connected with a  photon
collider based on the TESLA \EPEM\ linear collider are discussed. TESLA 
is the superconducting linear collider on the c.m.s. energy $2E_0 = 500$ GeV,
with the possibility of upgrading up to 800 GeV. In comparison with other
projects based on ``normal'' conducting structures
 TESLA has several advantages:
higher RF efficiency and correspondingly higher possible luminosity;
much larger distance between bunches which makes readout and
background problems much easier. Larger aperture of the accelerating
structure (and correspondingly smaller wake fields) allows bunches
with very small emittances to be accelerated without emittance dilution.
All this makes the TESLA project very attractive.  Many European high-energy
physicists  consider  TESLA as the next after LHC  project.
Now intensive work is going on both on accelerator, physics and 
detector with the intent to submit Technical Design Report (TDR) in Spring 
2001. 

Beside \EPEM\ collisions in TESLA project a second interaction
region for \GG\ and \GE\ collision is foreseen~\cite{Brinkmann98}.  It
is very important to include in the basic design of the collider
all that is necessary for a photon collider: special interaction region
with the crab crossing, minimization of beam emittances and their
preservation along the LC, space for lasers and laser optics, beam
dump etc. This work is underway. Detector for \GG, \GE\ collisions may
be very similar to that for \EPEM\ collision, though with some
complication connected with the focusing mirrors optics which should be
placed inside the detector.  The most important key element of the
photon collider is a laser system with very unique parameters.
Development of the laser system is currently the highest priority: ``To be
or not to be'' for photon colliders will be determined by the success of this
laser R\&D.

Below I will briefly discuss the following essential topics:
\begin{enumerate}

\item physics motivation;

\item possible parameters of the photon collider at TESLA;

\item lasers, optics;

\item backgrounds and some other issues.
\end{enumerate}

\section{Physics}

Physics motivation for photon colliders is a very important issue. The next
linear \EPEM\ collider should give some new information after LHC,
photon colliders should give something new (or better accuracy) in
addition to \EPEM. In general, physics in \EPEM\ and \GG, \GE\ collisions 
is quite similar because the same particles can be produced.
However, it is always better to study new phenomena in various
reactions because they give complementary information.  Some
phenomena can best be studied at photon colliders due to
better accuracy or larger accessible masses.

The second aspect important for physics motivation is the luminosity.
Typical luminosity distribution in \GG\ and \GE\ collisions has a high
energy peak and some low energy part (see the next section).  This
peak in \GG\ collisions has a width at half  maximum of about 15\%. The
photons in the peak can have a high degree of polarization.  In the next
section it will be shown that in the current TESLA designs the \GG\ 
luminosity in the high energy peak  can be
up to 30-40\% of the \EPEM\ luminosity at the same beam energy.
\subsection{Higgs boson}
The present Standard Model (SM) assumes existence of a very unique
particle, the Higgs boson, which is thought to be responsible for the
origin of particle masses. It has not been found yet, but from
existing experimental information it follows that, if it exists, its
mass is higher then 112 GeV (LEP200) and is below
200 GeV~\cite{Grutu}, i.e.  lays in the region of the next linear
colliders.  In the simplest extensions to the SM the Higgs sector
consists of five physics states: $h^0,H^0,A^0$ and $H^{\pm}$. All
these particles can be studied at photon colliders and some
characteristics can be measured better than in \EPEM\ collisions.

\begin{figure}[!htb]
\centering
\vspace*{-1.0cm} 
\hspace*{0cm} \epsfig{file=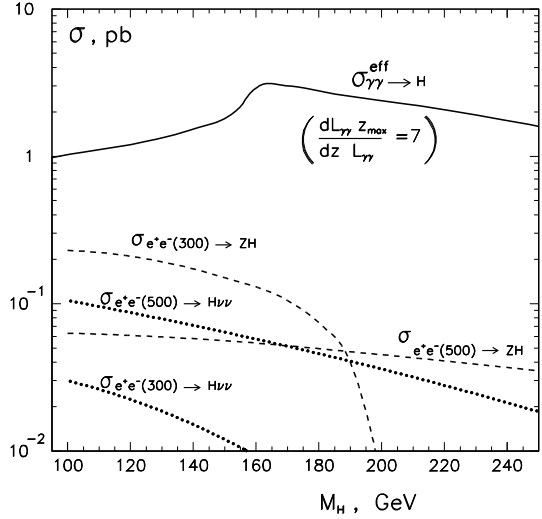,width=10cm,angle=0} 
\vspace*{-0.5cm} 
\caption{Cross sections for the Higgs in \GG\ and
 \EPEM\ collisions in the Standard Model.}
\vspace{2mm}
\label{cross}
\end{figure}

The process $\gamma\gamma \to H$ goes via the loop with heavy virtual charged
particles and its cross section is very sensitive to the contribution 
of particles with masses far beyond the energies covered by present and planned
accelerators~\cite{Okun}.  If the Higgs is light enough, its width is much less
than the energy spread in \GG\ collisions.   The ``effective'' cross section 
with account of the energy spread is presented 
in Fig.\ref{cross}~\cite{ee97}.
Here \LGG\ is defined as the \GG\ luminosity at the high energy peak
of the luminosity spectrum. For comparison, the cross sections of the
Higgs production in \EPEM\ collisions are shown.  We see that for
$M_H=$ 120--250 GeV the effective cross section in \GG\ collisions is
larger than the sum of the cross sections in \EPEM\ collisions by a
factor of about 8.5--30. The Higgs boson at photon colliders can be
detected as a peak in the invariant mass distribution or(and) can be
searched for by scanning the energy  using the sharp high energy edge of
the luminosity distribution~\cite{ee97}. For the integrated luminosity
50 fb$^{-1}$ (in the peak) the number of produced Higgs will be
50--150 thousands (depending on the mass).

At $M < 150$ GeV the Higgs decays mainly to b-quarks. The cross
section of the process $\gamma\gamma\to H \to b\bar{b}$ is
proportional to $\Gamma_{\GG}(H)\times Br(H\to b\bar{b}$). The
branching ratio $Br(H\to b\bar{b})$ can be measured with a high
precision in \EPEM\ collisions in the process with the "tagged" Higgs
production: $\EPEM\to ZH$~\cite{Bataglia}. As a result, one can
measure the $\Gamma_{\GG}(H)$ width at photon colliders with an
accuracy better than 2--3\%~\cite{Rembold99,Melles,Rembold00}.  This
is sufficient for distinguishing between Higgs models \cite{GinzKrav}.

Moreover in the models with several neutral Higgs bosons, heavy H$^0$
and A$^0$ bosons have almost equal masses and for wide range of
parameters are produced in \EPEM\ collisions only in associated
production $\EPEM\to HA$ \cite{Accomando}, while in \GG\ collision
they can be produced singly with sufficiently high cross
section~\cite{Muhlleitner}. Correspondingly, in \GG\ collisions one
can produce Higgs bosons with about 1.5 times higher masses.
\subsection {Charge pair production}
The second example is the charged pair production. It could be
$W^+W^-$ or $t\bar{t}$ pairs or some new, for instance, charged Higgs
bosons or supersymmetric particles.  Cross sections for the production
of charged scalar, lepton, WW pairs in \GG\ collisions are larger than
those in \EPEM\ collisions by a factor of approximately 5--20.
The corresponding graphs can be found elsewhere~\cite{TEL95,Brinkmann98}.

\begin{figure}[!thb]
\centering
\vspace*{-0.5cm}
\hspace*{-0.cm} \epsfig{file=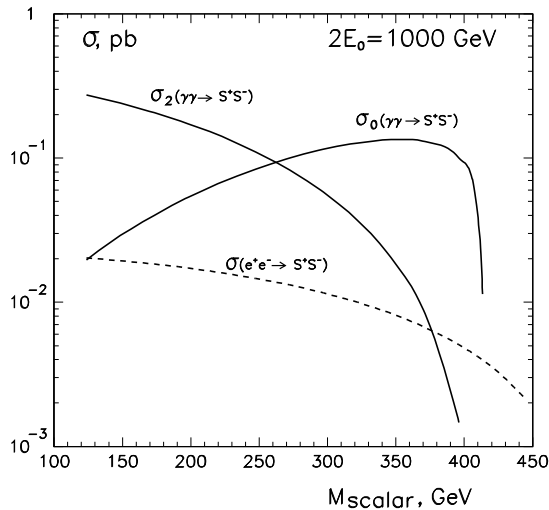,width=10cm}
\vspace*{-0.7cm}
\caption{ Cross sections for charged scalars production in \EPEM\ and
\GG\ collisions at $2E_0$ = 1 TeV collider (in \GG\ collision
$W_{max}\approx 0.82$ TeV, $x=4.6$); $\sigma_0$ and $\sigma_2$ correspond to
the total \GG\ helicity 0 and 2.}
\vspace{-0mm}
\label{crossel}
\end{figure}

For scalar particle, the cross section in \EPEM\ and \GG\ 
collisions is presented in Fig.\ref{crossel}~\cite{Tfrei}. One can see
that the cross section in collisions of polarized photons for large
masses (near the threshold) is higher than that in \EPEM\ collisions
by a factor of 20.  In addition, near the threshold the cross section in
the \GG\ collisions is very sharp, $\propto \beta$ (while in \EPEM\ it
contains a factor $\beta^3$); this is useful for measurements of
particle masses.

Note, that in \EPEM\ collisions two charged pairs are produced both via
annihilation diagram with virtual $\gamma$, $Z$ and also via
exchange diagrams where some new particles can give contributions,
while in \GG\ collisions it is pure a QED process which allows the
charge of produced particles to be measured unambiguously.  This is a
good example of complementarity in the study of the same particles in
different types of collisions.
\subsection{Accessible masses}
In \GE\ collisions, charged particle with a mass higher than that in
\EPEM\ collisions can be produced (a heavy charged particle plus a
light neutral); for example, supersymmetric charged particle plus
neutralino or new W boson and neutrino.  Also, \GG\ collisions  provide
higher accessible masses for particles which are produced as a single
resonance in \GG\ collisions (such as neutral Higgs bosons).

\subsection{Search for anomalous interactions}

Precise measurement of cross sections allow the observation of effects of
anomalous interactions. The process $\GG\to$ WW has large cross
section (about 80 pb) and it is one of most sensitive processes for
a search for a new physics (if no new particles are observed directly).
The vertex $\gamma WW$ can be studied much better than in \EPEM\ 
collisions because in the latter case the cross section is much smaller
and this vertex gives only 10\% contribution to the total cross
section. The two factors together give about 40 times difference in the cross
sections~\cite{Ginz2000}.  Besides that, in \GG\ collisions the $\gamma\gamma
WW$ vertex can be studied.

\subsection{Quantum gravity effects with Extra Dimensions}

This new theory~\cite{Arkani} suggests a possible explanation of why
gravitation forces are so weak in comparison with electroweak forces.
It is suggested that the gravitation constant is equal to the electroweak but
in a space with extra dimensions.  This extravagant theory can be
tested at photon colliders and a two times higher mass scale than in \EPEM\ 
collisions can be reached~\cite{RIZZO}.

Many other examples can be found in these proceedings. 

\section{Possible luminosities of \GG, \GE\ collisions at TESLA}

As it is well known in \EPEM\ collisions the luminosity is restricted
by beamstrahlung and beam instabilities. Due to the first effect the
beams should be very flat. In \GG\ collisions these effects are
absent, therefore one can use beams with much smaller cross section.
Limitations of the luminosity at photon colliders are discussed in
section 3.2. At present TESLA beam parameters the \GG\ luminosity is
determined only by the attainable geometric \EMEM\ luminosity.  Below
we consider first currently possible luminosity and then fundamental
limitations.

\subsection{Luminosities in the current design}

What luminosity can be obtained with the technology presently available? It
depends on emittances of electron beams. There are two methods of
production low-emittance electron beams: damping rings and
low-emittance RF-photo-guns (without damping rings).  The second
option is promising, but at this moment there are no such photo-guns
producing polarized electron beams~\cite{Ferrario}.  Polarization of
electron beams is very desirable for photon colliders~\cite{GKST84},
because: a) it increases the luminosity in the high energy peak by a
factor of 3--4; b) polarization characteristics of high energy photons
are better (broader part of spectrum have high degree of
polarization).  So, there is only one choice now --- damping rings.

Especially for a photon collider the possibility of decreasing the
beam emittances at the TESLA damping ring has been studied and
preliminary results were reported at the workshop~\cite{Decking}.
After some additional study it was found that the horizontal emittance
can be reduced by a factor of 4 compared to the previous design. Now
the normalized horizontal emittance is $\ENX\ = 2.5\times 10^{-6}$ m.

The luminosity depends also on $\beta$-functions at the interaction
points: $L \propto 1/\sqrt{\beta_x \beta_y}$.  The vertical $\beta_y$
is usually chosen close to the bunch length $\sigma_z$ (this is in
design for \EPEM\ collisions and can be done for \GG\ collisions as
well).  Some questions remain on the minimum horizontal
$\beta$-function. For \EPEM\ collisions, $\beta_x \sim 15$ mm which is
much larger than the bunch length $\sigma_z = 0.3$ mm, because beams
in \EPEM\ collisions must be flat to reduce beamstrahlung. In \GG\
collisions, $\beta_x$ could be about 1 mm (or even somewhat smaller).
There are two fundamental limitations: the beam length and the Oide
effects~\cite{Oide} (radiation in final quads), the latter is not
important for considered beam parameters. There is also a cetrain
problem with the angular spread of the synchrotron radiation emitted
in the final quads. But, for the photon collider the crab-crossing
scheme will be used and in this case there is sufficient clearence 
for the removal of the disrupted beams and synchrotron radiation.

Very preliminary studies of the existing scheme of the TESLA final
focus have shown~\cite{Walker} that chromo-geometric abberations
dominate at $\beta_x$ less than about 6 mm.  However, this value is
not fundamental limitation and it is very likely that after further
study and optimization a better solution will be found. Recently, at
SLAC a new scheme for the final focus system has been
proposed~\cite{Sery}.  The first check without optimization has
shown~\cite{Sery1} that with the new scheme one can obtain $\beta_x
\sim$ 1.5--2 mm with small abberations.  In this paper $\beta_x
\sim 1.5$ mm  is assumed.

Some uncertainties remain also for operation of  TESLA at low
energies. For the low mass Higgs (M$_H$ = 130 GeV) the required electron beam
energy lies between 79 GeV (for $x=4.6$, $\lambda=0.325$ \MKM\ ) and 100 GeV
(for $x=1.8$, $\lambda=1.06$ \MKM\ ). In this case TESLA should work
at lowered accelerating gradient or use bypass. If one use the same
electron bunches, the same beam-train structure and the repetition
rate, the luminosity will be proportional to the beam energy $E_0$
(because $L \propto 1/\sigma_x\sigma_y$ and $\sigma_i \propto
\sqrt{\EI}$ and $\EI\ = \ENI/\gamma$).

In principle, this loss of luminosity can be compensated by increase
of the repetition rate as $f \propto 1/E_0$, it this case the RF power
(for linac) is constant.  However, for the present design of the TESLA
damping ring, the repetition rate may be increased at most by a factor
of 2. Further decrease of the damping time is possible but at
additional cost (wigglers, RF-power).

Other problem of working at low gradients is the beam loading problem
(RF efficiency).  Its adjustment requires change of a coupler
position, which is technically very difficult or even impossible.  

Due to these uncertainties TESLA accelerator physicists have
recommended to use the scaling $L \propto E_0$ (bypass solution). However, if
the low mass Higgs is  found, some solutions for  increasing
luminosity at low energies can be found. For this paper I assume the
same beam parameters for all energies, that gives $L \propto E_0$.

The resulting parameters of the photon collider at TESLA for $2E_0=500$
GeV and H(130) are presented in Table 1. It is assumed that electron 
beams have 85\% longitudinal polarization and laser photons have 100\% circular
polarization. The thickness of laser target is one collision length 
($k^2 \approx 0.4$).  The conversion point (CP) is situated at the distance
$b=\gamma\sigma_y$ from the interaction point. 

\begin{table}[!hbtp]
\caption{Parameters of  the \GG\ collider based on TESLA. 
Left column for $2E_0=500$ GeV, next two columns for Higgs with $M_h=130$ GeV, 
two options.}
\vspace{5mm}
{\renewcommand{\arraystretch}{1.2}
\begin{center}
\begin{tabular}{l c c c} \hline
 & $2E_0=500$ & $2E_0=200$ & $2E_0=158$ \\
 & $x=4.6$ & $x=1.8$ & $x= 4.6$ \\  \hline 
$N/10^{10}$& 2 & 2 & 2  \\  
$\sigma_{z}$, mm& 0.3 & 0.3 & 0.3  \\  
$f_{rep}\times n_b$, kHz& 14.1 & 14.1 & 14.1  \\
$\gamma \epsilon_{x/y}/10^{-6}$,m$\cdot$rad & 2.5/0.03 & 2.5/0.03 & 
2.5/0.03 \\
$\beta_{x/y}$,mm at IP& 1.5/0.3 & 1.5/0.3 & 1.5/0.3 \\
$\sigma_{x/y}$,nm & 88/4.3 & 140/6.8 & 160/7.6  \\  
b, mm & 2.1 & 1.3 & 1.2 \\
\LEE(geom), $[10^{34}\, \CMS]$& 12 & 4.8 &  3.8 \\  
$\LGG (z>0.8z_{m,\GG\ }),[10^{34}\,\CMS] $ & 1.15 & 0.35 &  0.36  \\
$\LGE (z>0.8z_{m,\GE\ }),[10^{34}\, \CMS]$ & 0.97 & 0.31 & 0.27 \\
$\theta_{max}$, mrad & 9 & 8.5 & 16 \\ 
\end{tabular}
\end{center}
}
\label{table1}
\end{table}

Comparing the \GG\ luminosity with the \EPEM\ luminosity
($L_{\EPEM} = 3.4\times 10^{34}$ \CMS\ for $2E_0=500$ GeV \cite{Brinkmann99})
we see that for the same energy
\begin{equation}
\LGG(z>0.8z_m) \sim \frac{1}{3} L_{\EPEM}. 
\label{lgge+e-}
\end{equation}

The relation (\ref{lgge+e-}) is valid only for the considered beam
parameters. A more universal relation is (for $k^2=0.4$)
\begin{equation}
\LGG(z>0.8z_m) \sim 0.1\LEE(geom).   
\label{lgge-e-}
\end{equation}

For example, as we have seen (Fig.\ref{crossel}) the cross section for
production of $H^+H^-$ pairs in collisions of polarized photons is higher
than that in \EPEM\ collisions by a factor of 20 (not far from the
threshold); this means 8 times higher production rate for the
luminosities given above.

For the Higgs the production rate  is proportional to $dL_0/dW$ at $W=M_H$.
For the considered cases, $M_H =130$ GeV,  $x=4.6$ and $x=1.8$, $dL_0/dW =  
1.65\times 10^{32}/\GEV\ $ and $1.5\times 10^{32}/\GEV,$ respectively, so that
the coefficient in Fig.\ref{cross} characterizing the width of the peak is
about 5.9 and 5.5 (instead of 7). 

Using these luminosities and Fig.~\ref{cross} one can find that the
rate of production of the SM Higgs boson with M$_H$=130(160) GeV in
\GG\ collisions is 0.9(3) of that in \EPEM\ collisions at $2E_0 = 500$ GeV
(both reactions, ZH and H$\nu\nu$). 

\begin{figure}[!htb]
\centering
\vspace*{-2.0cm} 
\epsfig{file=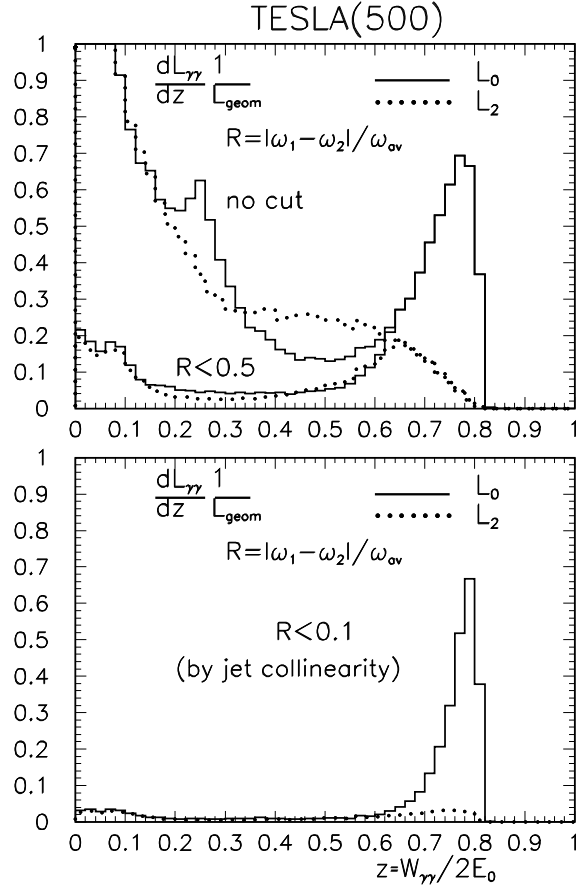,width=11cm,angle=0} 
\vspace{-2.0cm} 
\caption{\GG\ luminosity spectra at TESLA(500) with various cuts 
  on longitudinal momentum. Solid line for total helicity of two
  photons 0 and dotted line for total helicity 2.  See comments in the
  text.}
\vspace{0cm} 
\label{Ldist250}
\end{figure}

   The normalized \GG\ luminosity spectra for $2E_0 = 500$ GeV are
   shown in Fig.\ref{Ldist250}.
\begin{figure}[!htb]
\centering
\vspace*{-0.5cm} 
\epsfig{file=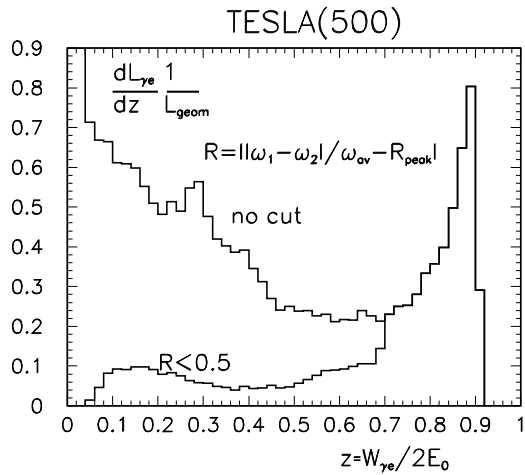,width=10cm,angle=0} 
\vspace{-1.2cm} 
\caption{Normalized \GE\ luminosity spectra at TESLA(500) for parameters
given in Table~\ref{table1}. See comments in the text.}
\vspace{0cm} 
\label{Ldist250e}
\end{figure} 
The luminosity spectrum is decomposed into two parts, with the total
helicity of two photons 0 and 2. We see that in the high energy part
of the luminosity spectra photons have a high degree of polarization.
In addition to the high energy peak, there is a factor 5--8 larger low
energy luminosity. It is produced mainly by photons after multiple
Compton scattering and beamstrahlung photons. These events have a
large boost and can be easily distinguished from the central high
energy events.  In the same Fig.\ref{Ldist250} you can see the same
spectrum with an additional cut on the longitudinal momentum of the
produced system which suppresses low energy luminosity to a low level.
For two jet events ($H\to b\bar b,\; \tau\tau$, for example) one can
restrict the longitudinal momentum using the acollinearity angle
between jets. The resulting energy spread of collisions will be
about 7.5 \% , see Fig.\ref{Ldist250} (lower).

The high energy part of the \GG\ luminosity spectrum is almost
independent on collision effects at the IP (beamstrahlung and multiple
Compton scattering). For theoretical studies one can obtain it with
sufficient accuracy by convolution of the Compton
function~\cite{GKST84}. Recently, a simple analytical formula for the
Compton function has been obtained~\cite{Galynskii} which takes into
account nonlinear effects in the conversion region for small enough
values of $\xi^2$. At photon colliders it is about 0.2--0.3 in the
center of the laser focus. In the simulation one has also to take into
account variation of $\xi^2$ in the conversion region. As a good
approximation one can generate "t" distributed by the Gaussian law and
then calculate $\xi^2=\xi^2_0\exp{(-t^2/2)}$.

The normalized \GE\ luminosity spectra for $2E_0=500$ GeV are shown in
Fig.\ref{Ldist250e}. Again, beside the high energy peak there is a
several times larger \GE\ luminosity at low invariant masses.  Note,
that \GE\ luminosity in the high energy peak is not a simple geometric
characteristic of the Compton scattering (as it is in \GG\ 
collisions). For the considered case it is suppressed by a factor of 2--3,
mainly due to repulsion of the electron beams and beamstrahlung.  The
suppression factor depends strongly on beam parameters. Note, that for
a special experiment on QCD study at low luminosity one can increase the
distance between the conversion and interaction points and obtain
very monochromatic \GE\ luminosity spectrum with a small low
energy background. This is due to the effective transverse magnetic
field in the detector solenoid which acts on the beam in the case
the crab crossing collisions.

\begin{figure}[!htb]
\centering
\vspace*{-1.5cm}
\hspace*{-0.0cm} \epsfig{file=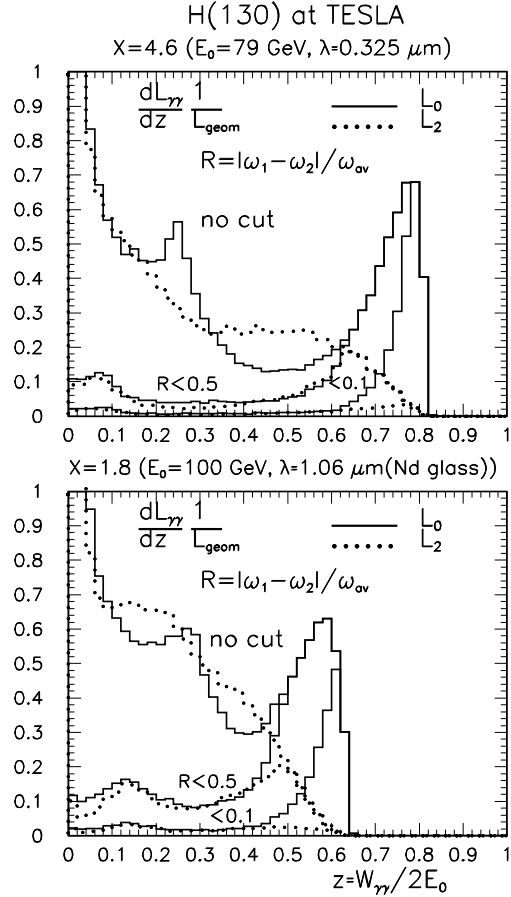,width=10.cm,angle=0}
\vspace*{-1.5cm}
\caption{\GG\ luminosity spectra at the photon collider
  for study of the Higgs with the mass $M_H=130$ GeV; upper figure for
  $x=4.6$ and lower one for $x=1.8$ (the same laser as for $2E_0=500$
  GeV). See also Table~\ref{table1}.}  
\vspace{-0.cm}
\label{Higgs130}
\end{figure}

The luminosity distributions for the case of Higgs (130 GeV) are
presented in Fig.\ref{Higgs130}. All parameters for this figure are
given in Table \ref{table1}.  Two cases are considered: $E_0=79$ GeV, $x
\sim 4.6$ ($\lambda$ = 0.325 \MKM\ ) and $E_0=100$ GeV,  $x=1.8$
($\lambda$ = 1.06 \MKM, which corresponds to  wave lengths of the
most power solid state lasers).

Beside the convenience of using the same laser with $\lambda \sim$ 1 \MKM\ 
for all energies there are two other advantages of the latter case. For
measurement of the Higgs bosons CP parity a linear polarization of
high energy photons is required~\cite{Gunion,Kramer}. Maximum
value of the linear polarization is $l_{\gamma} = 2/(1+x+(1+x)^{-1})$
\cite{GKST84}, it is 63.3 \% for $x=1.8$ and only 33.5 \% for $x=4.8$.
Second advantage of $x=1.8$ are smaller disruption angles of the beams
after collisions, see table \ref{table1}. The disruption angle and
the angular size of the first quadrupole determine the value of the
crab crossing angle (Fig.\ref{ggcol}). In the present design $\alpha_c=34$
mrad (angle between electron beams). It may be not sufficient for
collisions of 80 GeV electron beams and $x=4.6$. One can use here $\lambda \sim
1$ \MKM\ or $\lambda \sim 0.5$ \MKM\ (doubled frequency).

Several other important accelerator aspects of the photon collider
at TESLA are discussed in N.~Walker's talk at this workshop~\cite{Walker}.

\subsection{Ultimate \GG\ luminosity}
  
  There is only one collision effect restricting the \GG\ luminosity,
  that is a process of coherent pair creation when the high energy
  photon is converted into an \EPEM\ pair in the strong field of the
  opposing electron beam~\cite{CHTEL,TEL90,TEL95}. In
  \GE\ collisions, beside coherent pair creation, two other effects
  are important: beamstrahlung (electrons of the ``main'' electron
  beam radiate in the field of the electron beam used for photon
  production) and repulsion of beams.

\begin{figure}[!htb]
\centering
\vspace*{-0.8cm} 
\hspace*{-0.7cm} \epsfig{file=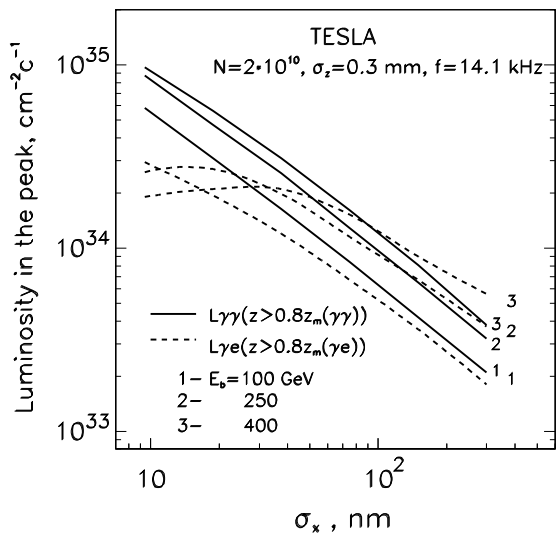,width=9.5cm,angle=0} 
\vspace*{-1.3cm} 
\caption{Dependence of \GG\ and \GE\ luminosities in the high energy
peak on the horizontal beam size for TESLA  at various
energies. See also comments in the text.}
\vspace{0mm}
\label{sigmax}
\vspace{1mm}
\end{figure}

 Below are results of  simulations with the
code which takes into account all main processes in beam-beam
interactions~\cite{TEL95}.  Fig.\ref{sigmax} shows dependence of the
\GG\ (solid curves) and \GE\ (dashed curves) luminosities on the
horizontal beam size for TESLA beam parameters at several energies.
  The horizontal beam size was varied by change of horizontal 
beam emittance keeping the horizontal beta function at the IP 
constant and equal to 1.5 mm.

One can see that all curves for \GG\ luminosity follow their natural
behavior: $L\propto 1/\sigma_x$ (values $\sigma_x < 10$ nm are not
considered because too small horizontal sizes may have problems with
crab-crossing).  Note that in \EPEM\ collisions $\sigma_x\sim 500$ nm.
In \GE\ collisions the luminosity is lower than in \GG\ collisions due
to the displacement of the electron beam during the beam collision and
beamstrahlung.  

So, we can conclude that for \GG\ collisions at TESLA
one can use beams with a horizontal beam size down to 10 nm (maybe
even smaller) which is much smaller than that in \EPEM\ collisions.
Note, that the vertical beam size could also be additionally decreased
by a factor of two (for even smaller electron beam size the effective
photon beam size will be determined by the Compton scattering
contribution). As a result, the \GG\ luminosity in the high energy
peak can be, in principle, several times higher than the \EPEM\
luminosity.

Production of the polarized electron beams with emittances lower than
those possible with damping rings is a challenging problem.  There is
one method, laser cooling~\cite{TSB1,Monter,Tlasv1} which allows, in
principle, the required emittances to be reached.  However this method
requires laser power  one order of magnitude higher than is
needed for $e \to \gamma$ conversion. This is not excluded, but since
many years of R\&D would be required, it should be considered as a
second stage of the photon collider, maybe for the Higgs factory.

\section{Lasers, optics}

A key element of photon colliders is a powerful laser system which is
used for e$\to\gamma$ conversion.  Lasers with the required flash
energies (several J) and pulse duration $\sim$ 1 ps already exist and
are used in several laboratories. The main problem is the high
repetition rate, about 10--15 kHz, with a pulse structure repeating
the time structure of the electron bunches.  

A very promising way to overcome this problem at TESLA is an
``external optical cavity'' approach which allows a considerable
reduction of the required peak and average laser power~\cite{Tfrei}.
Technical aspects of this approach are considered by I.~Will and his
colleagues at this Workshop~\cite{Will}.  

Another possible solution is a one-pass free electron laser. Recently,
first such laser has been successfully commissioned at
DESY~\cite{DFEL}.  This option is described by M.Yurkov and
colleagues~\cite{Yurkov} at this Workshop.  

However, the most attractive and reliable solution at this moment is
an ``optical storage ring'' with a diode pumped solid state laser
injector which, it seems, can be build already now. This new approach
can be considered as a base-line solution for the TESLA photon
collider.

\subsection{Requirements for the laser, wave length, flash energy}   
The processes in the conversion region, i.e. Compton scattering and
several other important phenomena, have been considered in detail in
papers~\cite{GKST83,GKST84,TEL90,TEL95,Monter,Serbo} and references
therein. Laser parameters important for this task are: laser flash
energy, duration of laser pulse, wave length and repetition rate.  The
requirement for the wave length was considered in the introduction.
The optimum value for TESLA(500) is $\lambda \sim$ 1 \MKM.

In the calculation of the required flash energy one has to take into
account the natural ``diffraction'' emittance of the laser
beam~\cite{GKST83} and nonlinear effects in the Compton scattering.
The nonlinear effects are characterized by the parameter $\xi^2 =
e^2\bar{B^2}\hbar^2/(m^2c^2\omega_0^2)$ \cite{Berestetskii}
($\omega_0$ is the energy of laser photons).  Due to nonlinear effects
the high energy peak in the Compton spectrum is shifted to lower
energies: $\omega_m/E_0 = x/(x+1+\xi^2)$, this leads also to
broadening of the high energy edge of the luminosity spectrum. Therefore
$\xi^2<0.2-0.3$ is required~\cite{TEL95}.

\begin{figure}[!htb]
\centering
\vspace{-1.cm}
\epsfig{file=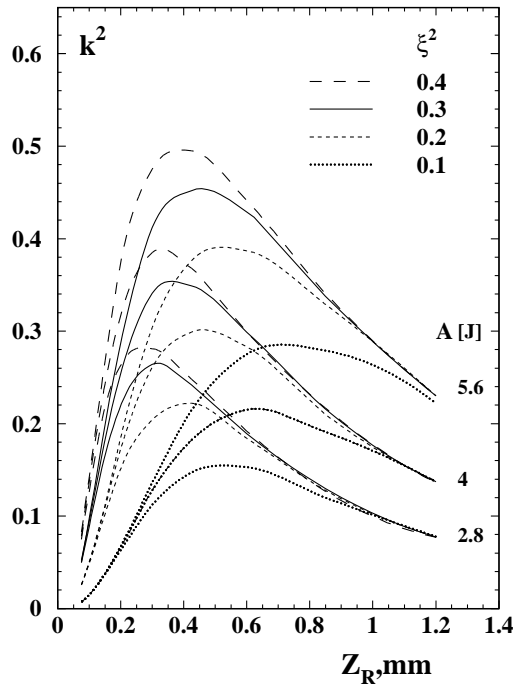,width=8cm,angle=0}
\vspace{-1.cm}
\caption{Square of the conversion probability (proportional to the \GG\ 
  luminosity) as a function of the Rayleigh length for various
  parameters $\xi^2$ and laser flash energies, $x=4.8, \, \lambda =1.06$
  \MKM\ are assumed. The mirror system is situated outside the
  electron beam trajectories (collision angle $\theta =
  2\sigma_{L,x^{\prime}}$). The crab crossing angle 30 mrad is taken
  into account.}
\label{k2}
\end{figure}

The result of MC simulation of $k^2$ ($k$ is the conversion
coefficient) for the electron bunch length $\sigma_z= 0.3$ mm (TESLA
project), $\lambda=1.06$ \MKM, $x=4.8$ as a function of the Rayleigh
length ($\beta$-function of the laser beam at a focal point) for
various flash energies and values of the parameter $\xi^2$ (in the
center of the laser bunch) are shown in Fig.\ref{k2}.

It was assumed that the angle between the laser optical axis and the
electron beam line is $\theta = 2\sigma_{L,x^{\prime}}$
($\sigma_{L,x^{\prime}}$ is the angular divergence of the laser beam
in the conversion region), so that the mirror system is situated
outside the electron beam trajectories. One conversion length
corresponds to $k^{2} = (1- e^{-1})^2 \approx 0.4$. One can see that
$k^2=0.4$ at $\xi^2=0.3$ can be achieved with the minimum flash energy
$A=5$ J. The optimum value of the Rayleigh length is about 0.35 mm.

The r.m.s. duration of the laser pulse can be expressed via other
laser parameters defined above~\cite{Brinkmann98}
\begin{equation}
\sigma_{L,z} = \frac{4r_e\lambda A}{(2\pi)^{3/2}mc^2 \xi^2 Z_R} =
9.3\times 10^{-5} \frac{A[\mbox{J}]}{Z_R[\CM]\xi^2}\;\mbox{cm},
\end{equation}
where the last equality corresponds to $\lambda=1.06$ \MKM. For $\xi^2
= 0.3$ and   $Z_r = 0.35$ mm the optimum length of the laser bunch
is $\sigma_{L,z}=0.44$ mm or 1.5 ps.

The angular size of the focusing mirror is taken to be equal to two
r.m.s. angular spread (in $x$ or $y$ directions) at the conversion
point~\cite{GKST83}
\begin{equation}
\Delta \theta_m = \pm 2 \sigma_{L,x^{\prime}} = \pm 2 
\sqrt{\frac{\lambda}{4\pi Z_R}} = \pm 0.031.
\label{thm}
\end{equation} 
The ratio of the focal distance to the mirror diameter ($f_\#$) is
equal to $1/(2\times 0.031) \approx 16$. For $L=2$ m, $d=12.5$ cm.  The
fluence ($A/$cm$^2$) in the center of the mirror is
$A/(2\pi\sigma_x^2)=8A/(\pi d^2) \sim 0.082$ J/cm$^2$, while the
damage threshold is in the range 0.7--2 J/cm$^2$~\cite{NLC}. In addition to the
final focusing mirror the optics inside the detector contains another
 mirror with about two times smaller diameter~\cite{Brinkmann98}
(see also Fig.\ref{loop}). The corresponding fluence on this mirror is
four times higher, about 0.3 J/cm$^2$. In the case of flattop laser
beams approximately similar beam sizes in focus are obtained when the
$f_\#$ of the flattop beam is about 0.7 times that of the Gaussian
beam~\cite{NLC}. The fluence on the mirrors will be about 4 times
smaller (due to larger area and uniform distribution).

In summary: the required laser wavelength is about 1 \MKM, flash
energy is 5 J, the repetition rate of about 14 kHz, so the average
power of the laser system should be about 70 kW each of two lasers.
Besides, at TESLA the laser should work only 0.5 \% of the time: one
train of 1 msec duration contains 3000 bunches, the repetition rate is
5 Hz. Non-uniformity of LC operation (the train structure) is a very
serious complication.

Below we will consider three variants of laser system: optical storage
ring, external optical cavity, free-electron laser. Present technique
of short pulse powerful solid state lasers is based on the following
modern technologies: pulsed-chirped techniques, diode pumping, adaptive
optics. Discussion of laser technologies for photon colliders can be
found elsewhere~\cite{NLC}.

\subsection{Optical storage ring}

\begin{figure}[p]
\centering
\vspace*{0.cm} 
\hspace*{-2.1cm} \epsfig{file=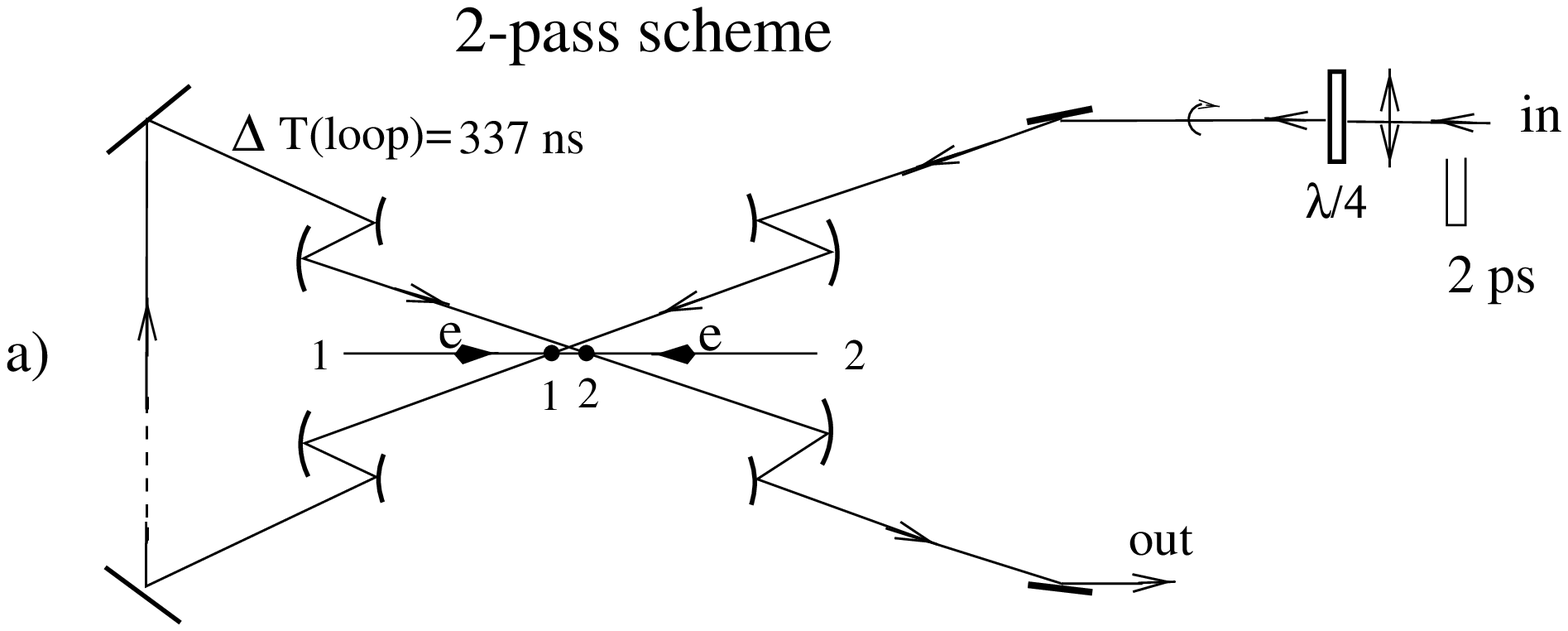,width=10.8cm,angle=0} 
\vspace*{0.3cm} 
\hspace*{-0.cm} \epsfig{file=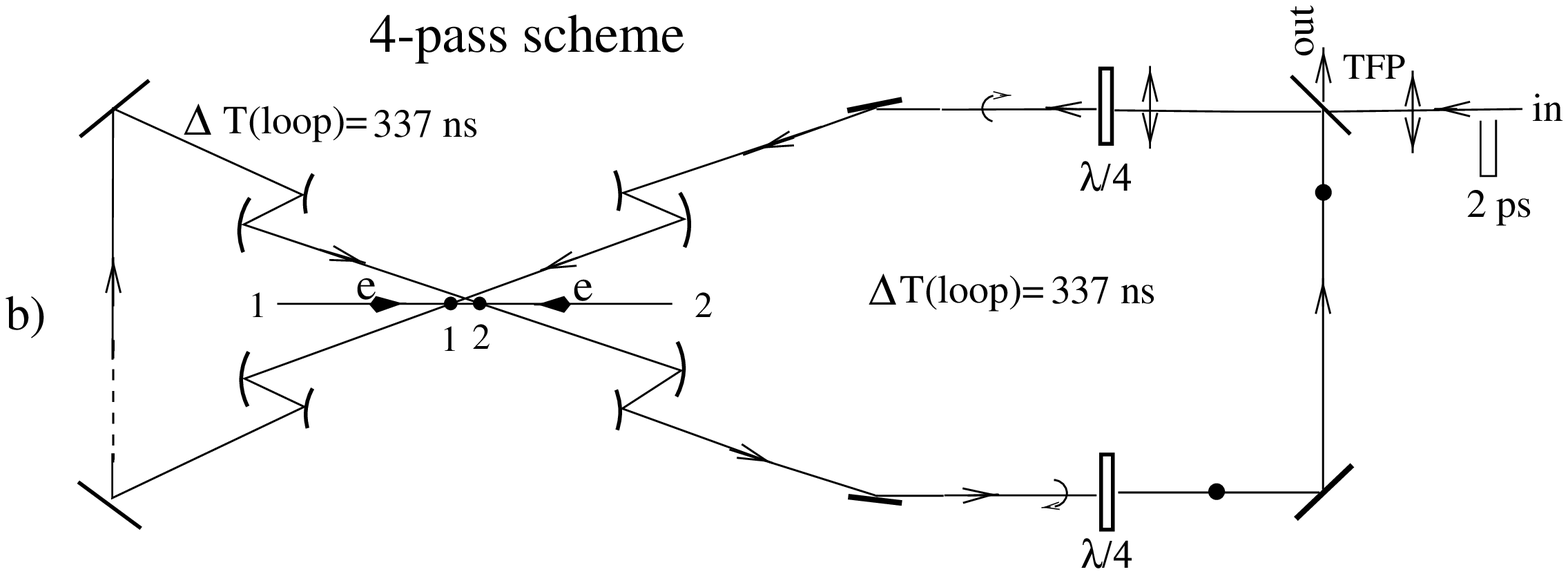,width=13.cm,angle=0} 
\vspace*{0.3cm}
\hspace*{-0.cm} \epsfig{file=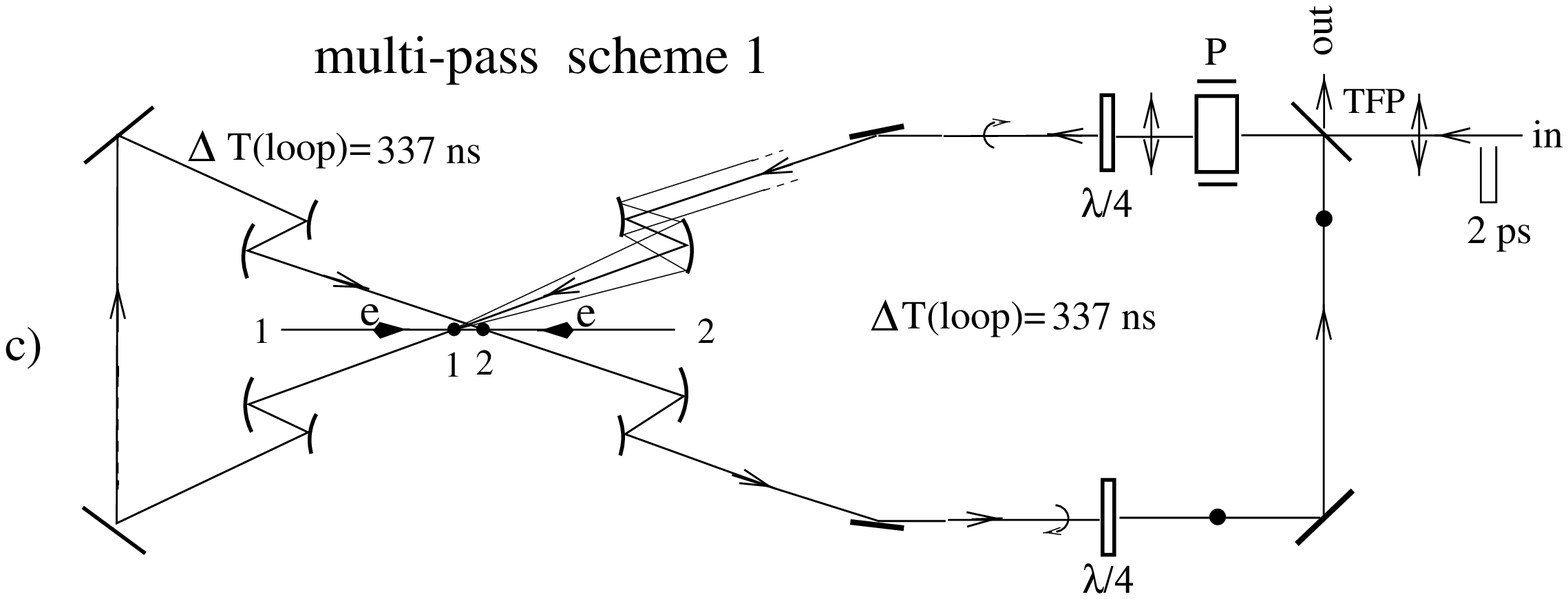,width=13.cm,angle=0} 
\vspace*{0.3cm}
\hspace*{-0.0cm} \epsfig{file=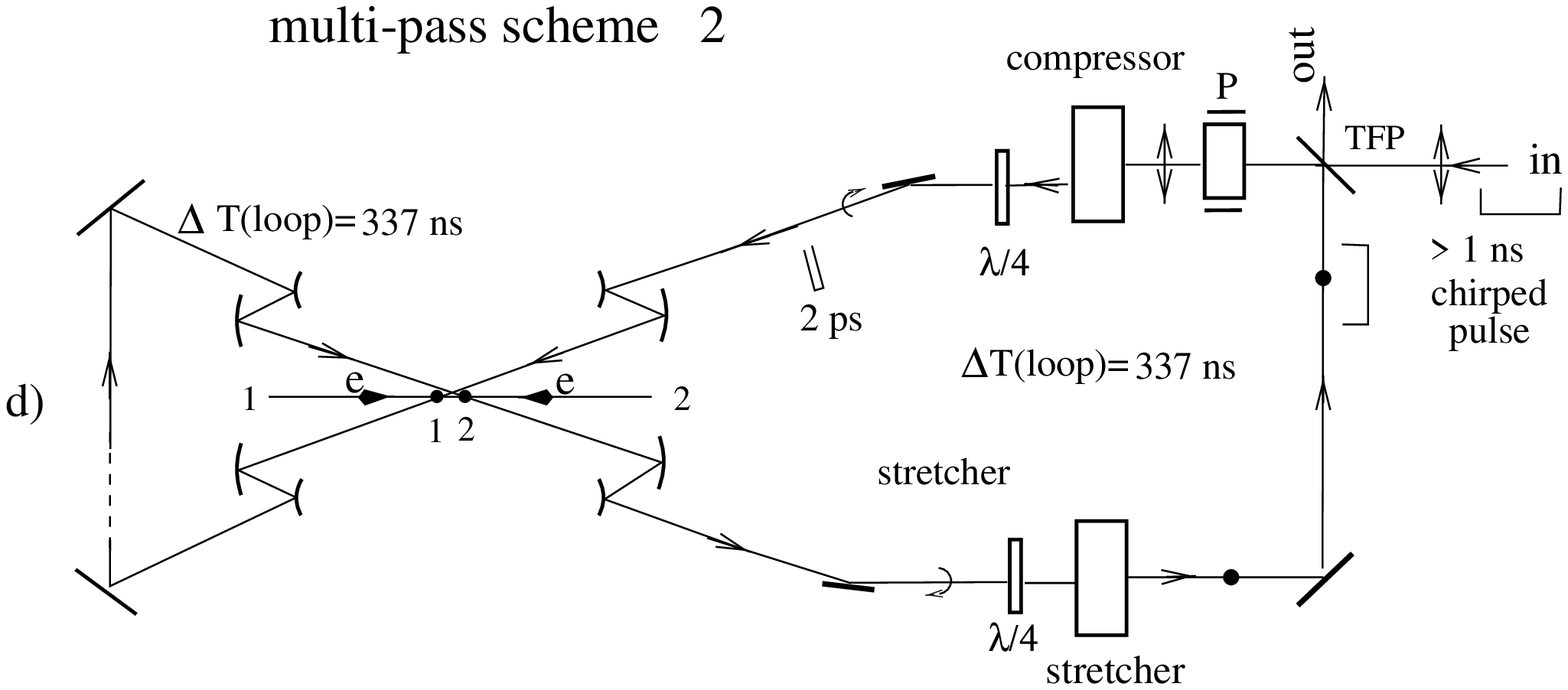,width=13.cm,angle=0} 
\vspace*{-0.0cm} 
\caption{Optical trap: a) 2-pass optics for $e \to \gamma$
conversions; b) 4-pass optics; c) optical storage ring without
stretching-compression; d) optical storage ring with
stretching-compression; P is a Pockels cell, TFP is a thin film
polarizer, thick dots and double arrows show the direction of
polarization.}  \vspace{0mm}
\label{loop}
\vspace{-0mm}
\end{figure}

To overcome the ``repetition rate'' problem it is quite natural to
consider a laser system where one laser bunch is used for e$\to
\gamma$ conversion many times. Indeed, 5 J laser flash contains about
$5\times 10^{19}$ laser photons and only $10^{10}-10^{11}$ photons are
knocked out in one collision with the electron bunch.
Below   two ways of multiple use of one laser pulse is considered. The
first approach is shown in Fig.~\ref{loop}. 
 
In Fig.\ref{loop}a the laser pulse is used twice for $e \to \gamma$
conversion.  After the collision with the electron beam
(number 1) the laser beam exits from the detector and after a 337 ns
loop (the interval between beam collisions at TESLA) returns back and
collides with the opposite electron beam (number 2). The second pass
does not need any special optical elements, only mirrors. This is a very
natural and simple solution. In this scheme  the laser system should generate
bunches with an interval of 337 ns.

In Fig.\ref{loop}b the laser pulse is used for conversion four times.
In this scheme one additional optical element is used, a thin film
polarizer (TFP), which is transparent for the light polarized in the
plane of the paper and reflects the light with the orthogonal
polarization.  Directions of the polarization during the first cycle
are shown in Fig.\ref{loop}b. After the first cycle the polarization is
perpendicular to the paper and the light is reflected from the TFP,
while after the second cycle the polarization will be again in the
plane of the paper and the laser pulse will escape the system via the TFP.
The laser bunches are emitted by the laser at 
an average interval of 2$\times$337 ns but not uniformly 
(337,3x337),(337,3x337), etc (see the next paragraph).

In Fig.\ref{loop}c the laser pulse is sent to the interaction region
where it is trapped in an optical storage ring.  This can be done
using Pockels cells (P), thin film polarizers (TFP) and 1/4-wavelength
plates ($\lambda/4$). Each bunch makes several (n) round trips and
then is removed from the ring.  All these tricks can be done by
switching one Pockels cell which can change the direction of linear
polarization by 90 degrees. The $\lambda/4$ plates are used for
obtaining circular polarization at the collision point.  For obtaining
linear polarization at the IP these plates should be replaced by 1/2
wavelength plates. A similar kind of optical trap was considered as
one of the options in the NLC Zero-design report~\cite{NLC}.  The
number of cycles is determined by the attenuation of the pulse and by
nonlinear effects in optical elements.  The latter problem is very
serious for TW laser pulses.  During one total loop each bunch is used
for conversion twice (see Fig.\ref{loop}c). The laser bunch is
collided first with electron beam 1 traveling to the right and after a
time equal to the interval between collisions (333 ns) it is collided
with beam 2 traveling to the left.  For arbitrary number of the round
trips, $n$, the laser pulse sequence is a sum of two uniform trains
with the interval between neighboring pulses in each train
\begin{equation}
\Delta T_t =2nT_0  
\label{DTt}
\end{equation}
and the trains are shifted on the time
\begin{equation}
\Delta T = k T_0, \;\;\;\; k=1,3, \ldots 2n-1.  
\label{Dt}
\end{equation}

In Fig.\ref{loop}d the laser pulse is trapped in the same way as in
Fig.9c, but to avoid the problems of non-linear effect (self-focusing)
in the optical elements, a laser pulse is compressed before the collision
with the electron bunch down to about 2--3 ps using grating pairs, it is
then stretched again up to the previous length of about 1 ns just before
passing through the optical elements.

Which system is better, \ref{loop}c or \ref{loop}d, is not clear
apriori. In the scheme (c) the number of cycles is limited by
nonlinear effects, in the scheme (d) additional attenuation is added
by the gratings used for compression and stretching. Optical companies
suggest gratings for powerful lasers with $R\sim 95$ \%. One round
trip needs four gratings, this means 20\% loss/trip. So, the maximum
number of trips is only about two. This present no advantages compared
to  the scheme \ref{loop}b which is simpler and also allows two
cycles though it is not excluded that gratings with higher reflectivity will
be available in future.

How large can the decrease of the laser energy per round trip be in
the scheme without bunch compressor-stretchers? The minimum number of
mirrors in the scheme is about 15--20. Reflectivity of multilayer
dielectric mirrors for large powers suggested by optical companies is
about 99.8\% or even better. So, the total loss/cycle is about
3--4\%. Let us add 1\% attenuation in the Pockels cell. Due to the
decrease of the laser flash energy the luminosity will vary from
collision to collision.  Calculation shows that for 1.3, 1.4, 1.5
times attenuation of the laser pulse energy before the pulse is
replaced, the \GG\ luminosity will only vary on 14, 17, 21 \%  (here we
assumed that on average the thickness of the laser target is one
collision length).  For 5\% loss/turn and {\it 6 round trips} the
attenuation is 1.35, which is still acceptable.

Let us consider the problem of nonlinear effects for the scheme \ref{loop}c.
The refractive index of the material depends on the beam intensity
\begin{equation}
 n = n_0 +n_2 I.
\end{equation}
This leads to self focusing of the laser beam~\cite{koechner}. There
are two kinds of self-focusing.  The first type is a self-focusing of
the beam as a whole. The second one is  self-focusing and amplification of
nonuniformities  which leads to break up of the beam into a large number
of filaments with the intensity exceeding the damage level.  Both
these effects are characterized by the parameter
B-integral~\cite{koechner,NLC}
\begin{equation}
B= \frac{2\pi}{\lambda} \int  \Delta n dl = 
\frac{2\pi}{\lambda} n_2 I_{peak} L 
\label{Bint}
\end{equation}

If the beam has a uniform cross section then nonlinear effects do not
lead to a change of the beam profile, while for the Gaussian like beam
$B \sim 1$ means that the self-focusing angle is approximately equal
to the diffraction divergence of the beam. This is not a problem, such
kind of distortions can be easily corrected using adaptive optics
(deformable mirrors).

The second effect is more severe. Even for uniform (in average)
distribution of the intensity over the aperture a small initial
perturbation $\delta{I}$ grows exponentially with a rate depending on
the spatial wave number.\footnote{Such instability exists when total
  beam power exceeds the threshold power for the medium, which is of
  the order of one MW~\cite{koechner}, that is 6 orders lower than the
  power needed for photon colliders} The maximum rate is given by the same
parameter $B$ \cite{koechner}
\begin{equation}
\delta{I} = \delta{I}_0 e^B.
\label{deltaI}
\end{equation}
This is confirmed experimentally. To avoid small-scale growth the parameter
$B$ should be smaller than 3--4~\cite{koechner,NLC}, in other words
\begin{equation}
I_{peak} < \frac{\lambda}{2n_2 L}
\end{equation}

Now we can get the relationship between the diameter and the maximum
thickness of the material.  For $A=5$ J, $\lambda = 1$ \MKM,
$\sigma_{L,z}$ = 1.5 ps and a typical value of $n_2 \sim 3\times
10^{-16}$ cm$^2$/W (I have not found in literature $n_2$ for KD$^*$P
used for Pockels cells) and flat-top beam we get
\begin{equation}
L[\mbox{cm}] < 0.1 S [\CM^2].
\end{equation}
For a beam diameter of 15 cm we obtain $L < 17$ cm.
For Gaussian beams the maximum thickness is about two times smaller.

What is L? Is it the maximum thickness of one optical element, the thickness 
integrated over the  loop or the previous one times the number of
round trips? In the scheme \ref{loop}c the dominant contribution to
the total thickness is given by the Pockels cell. After the Pockels cell one
can put a spatial filter and thus  suppress the growth of spikes. So,
$L$ in our case is the thickness of the Pockels cell and it does not
depend on the number of round trips. Moreover, our laser pulse is very
short, has a broad spectrum and the corresponding coherence length is
small, about $l_c \sim 4\pi\sigma_{L,z} \sim 0.5$ cm. The
instabilities upon a uniform high intensity background  developes due to
interference of fluctuation with the main power. However, this
coherence is lost after the coherence length. In my opinion, the
B-integral does not characterize the exponential growth of
irregularities once the coherence length is much lower than $L$. So,
it seems, that the problem of nonlinear effects in the scheme
\ref{loop}c is not dramatic.  Construction of a Pockels cell with an
aperture of about 10--15 cm and switching time 300 ns is not a very difficult.
Quarter- and half-wave plates can be made thin or even joined
with mirrors (retarding mirror).

In conclusion, it seems, that the optical scheme \ref{loop}c with
about 6 round trips (12 collisions with electron beams) is a very
attractive and realistic solution for the TESLA photon collider.

\begin{figure}[!htb]
\centering
\vspace*{-0.0cm} 
\hspace*{-0.4cm} \epsfig{file=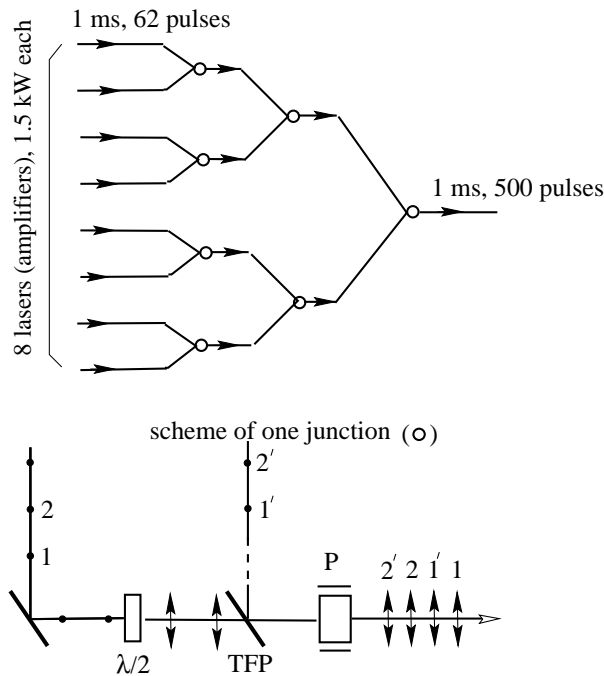,width=8cm,angle=0} 
\vspace*{-0.0cm} 
\caption{Merging of pulses from several lasers (amplifiers).}
\vspace{0mm}
\label{junction}
\vspace{-0mm}
\end{figure}

Now a few words on a laser system required for a such an optical
storage ring with 6 round trips.  Schematically it is shown in
Fig.\ref{junction}.  At the start (this is not shown) low-power laser
produces a train of 1 msec duration consisting of 500 chirped pulses
with duration several ns each.  Then these pulses are distributed
between 8 final amplifiers (see Fig.\ref{junction}).  Each of the 8
sub-trains have a duration of 1 ms and consist of 62 pulses.  After
amplification up to the energy 5 J in one pulse these sub-trains are
recombined to reproduce the initial time structure. The time spacing
between bunches in the resulting train is equal in average to 6
intervals between beam collisions in TESLA.

Due to the high average power the lasers should be based on diode
pumping. Diodes have much higher efficiency than flash lamps, about
$\epsilon \sim 25$ \% for single pulses.  For pulse train (as it is in
our case) it should be at least by factor of two higher. Besides that,
diodes are much more reliable. This technology is developed very
actively for other application, such as inertial fusion.

Main problem with diodes is their cost.  Present cost of diode lasers
is about 5\$ per Watt~\cite{Gronberg}.  Let us estimate the required
laser power. In the case of TESLA, the duration of the pulse train
T$_0$ = 1 msec is approximately equal to the storage time
($\tau\sim 1$ msec) of the most promising powerful lasers crystal, such as
Yb:S-FAP. Therefore,  the storage time does not help at
TESLA. The required power of the diode pumping is
\begin{equation}
P_{diode} = \frac{A(\mbox{flash})N(\mbox{bunches})}{\epsilon T_0} = 
\frac{5\;\mbox{J} \times 500}{0.5 \times 10^{-3}\;} = 5\; 
\mbox{MW}. 
\label{huge}
\end{equation}
Correspondingly, the cost of such diode system will be 25 M\$.  Here
we assumed 6-fold use of one laser bunch as described above.  We see
that even without Pockels cell (the scheme \ref{loop}b with two round
trips) the diode power is 15 MW and the cost of diodes is then 75
M\$,  1--2 percents of the LC cost.

Moreover, the Livermore laboratory is now working on a project of
inertial confinement fusion with a high repetition rate and
efficiency with the goal to build a power plant based on fusion. This
project is based on diode pumped lasers.  According to
ref.\cite{Payne} they are currently working on the ``integrated
research experiment'' for which ``the cost of diodes should be reduced
down to 0.5\$/Watt and the cost of diodes for fusion should be
0.07\$/Watt or less.'' Thus, the perspectives of diode pumped lasers for
photon colliders are very promising. With 1\$/Watt the cost of diode
is 5 M\$ for 6 round trips (with Pockels cell) and 15 M\$ for 2 round
trips without Pockels cell.

Average output power of all lasers in the scheme \ref{loop}c is about 12 kW,
1.5 kW for each laser. 

\subsection{``External" optical cavity}

One  problem with the optical storage ring at photon colliders is the
self-focusing in optical elements due to the very high laser pulse power. 
There is another  way to ``create'' a powerful laser pulse in
the optical ``trap'' without any material inside: laser pulse 
stacking in an ``external'' optical cavity~\cite{Tfrei}.

In short, the method is the following. Using the train of low energy
laser pulses one can create in the external passive cavity (with one
mirror having some small transparency) an optical pulse of the same
duration but with the energy by a factor of Q (cavity quality factor)
higher.  This pulse circulates many times in the cavity each time
colliding with electron bunches passing the center of the cavity. For
more details see ref.~\cite{Tfrei}. 
   
\begin{figure}[!htb]
\centering
\vspace*{-0.cm} 
\hspace*{-0.4cm} \epsfig{file=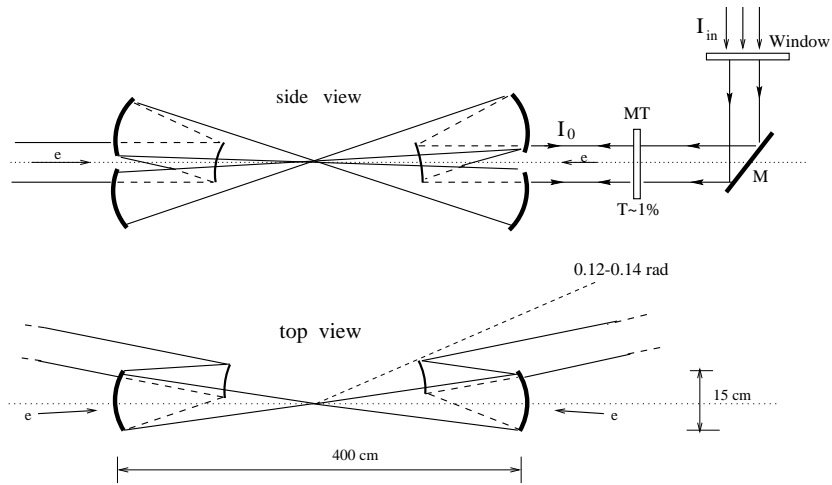,width=11cm,angle=0} 
\vspace*{-0.0cm} 
\caption{Principle scheme of ``external'' cavity  for $e \to \gamma$ 
  conversion. Laser beam coming periodically from the right
  semi-transperant mirror MT excites one cavity (includes left-down
  focusing mirror, right-up focusing mirror and the MT mirror. The
  second cavity (for conversion of the opposite electron beam) is
  pumped by laser light coming from the left (not shown) and includes
  the focusing mirrors left-up and right-down.}
\vspace{0mm}
\label{optics}
\vspace{-0mm}
\end{figure}

\begin{figure}[!htb]
\centering
\vspace*{0.3cm} 
\hspace*{-0.4cm} \epsfig{file=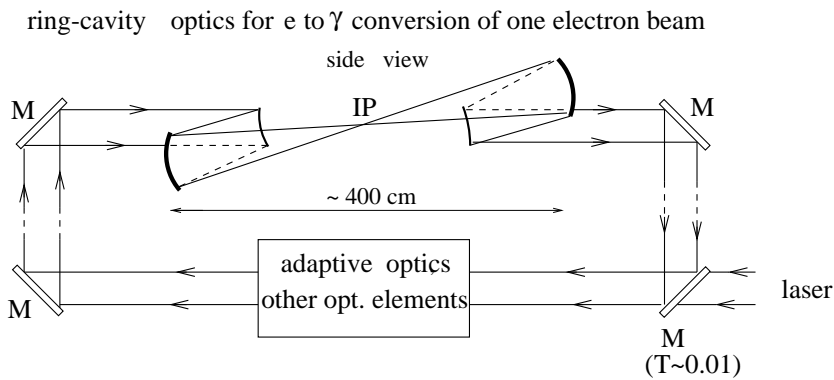,width=11cm,angle=0} 
\vspace*{-0.0cm} 
\caption{Ring type cavity. Only the cavity for one electron beam is shown.
Top view is quite similar to that in Fig.\ref{optics}.}
\vspace{0mm}
\label{ring}
\vspace{-0.3mm}
\end{figure}

A possible layout of the optics at the interaction region scheme is
shown in Fig.\ref{optics}. In this variant, there are two optical
cavities (one for each colliding electron beam) placed outside the
electron beams.
Such a system has minimum numbers of mirrors inside the detector.  One
of several possible problems in such linear cavity (pointed out by I.Will at
this workshop~\cite{Will}) is back-reflection, in a ring type cavity
this problem would be much easier. A possible scheme of such a
ring cavity for photon colliders is shown in Fig.\ref{ring} (only
some elements are shown). 
 
More detailed technical aspects of the external cavity approach were
discussed by I.Will and colleagues at this workshop~\cite{Will}.  Such
a cavity is operated already in their institute and a $Q \sim 100$
has been demonstrated. In their opinion, the laser system with the optical
cavity required for the photon collider can be made, though this task
is much more difficult due to the much higher powers.

Comparing the optical cavity approach with the ``optical trap''(or
``optical storage ring'') considered in previous subsection, I would
say that the optical cavity is very attractive but less studied. Here
all tolerances should be of the order $\lambda/(2\pi Q)$, while in the
optical trap they are much more relaxed, about $0.2\sigma_z/ Q$, 
difference 300 times for the same $Q$!

\subsection{Free electron lasers}

Anther approach for a laser for the TESLA photon collider is a free
electron laser~\cite{Yurkov}. Recently DESY has demonstrated operation
of a one pass FEL with the wave length 0.1 \MKM\ and further goal is 1
nm (this is included in TESLA proposal). Generation of 1 \MKM\ light is much
a easier task. However this task is also not simple because to
obtain the required flash energy a very large part of the electron
beam energy should be converted to light.

\subsection{Laser damage of optics}
\vspace{0.3cm}

The peak and average power in the laser system at the photon collider is
very large. The damage threshold for multilayer dielectric mirrors
depends on the pulse duration. The empirical scaling law is~\cite{koechner}
\begin{equation}
E_{th}[J/cm^2] \sim 10 \sqrt{t[ns]}
\end{equation} 
for pulse durations ranging from picoseconds to milliseconds.  At the
LLNL the damage threshold for 1.8 ps single pulses of 0.7 to 2
J/cm$^2$ have been observed on commercial multilayer
surfaces~\cite{NLC} with an average flux on the level of 3--5 kW/cm$^2$.

Comparing these numbers with the conditions at the TESLA photon
collider (5 J for 1.5 ps, 6000x5 J for 1 ms and 140 kW average power)
one finds that the average power requirements are most demanding. With
uniform distribution, the surface of the mirrors should be larger than
140/5 = 28 cm$^2$ and a factor of 2--3 larger for gaussian laser beams
with cut tails. So, the diameter of the laser beam on mirrors and
other surfacies should be larger than 10 cm.

\vspace{0.3cm}
{\bf Short summary on lasers and optical schemes}
\vspace{0.3cm}

We have considered 3 possible options of laser system for
TESLA photon collider:

\begin{enumerate} 
  
\item \underline{Optical trap (storage ring)} with about 8 diode
  pumped driving lasers (final amplifiers) with total average power
  about 12 kW.  Beams are merged to one train using Pockels cells and
  thin-film polarizers. Each laser pulse makes 6 round trips in the
  optical trap colliding 12 times with the electron beams. This can be
  done now: all technologies exist.

\item \underline{External optical cavity.} Very attractive solution, but
needs very high tolerances and mirror quality. Serious R\&D is required.

\item \underline{Free electron laser.} Is attractive due to variable
  wave length.  High electron to photon energy conversion efficiency
  required is a problem.

\end{enumerate}

\section{Experimentation issues}

Backgrounds at the TESLA photon collider were discussed in the TESLA
Conceptual design~\cite{TESLA}. More simulations are necessary with
detailed geometry and material. Here we consider only two geometric
questions connected with optics inside the detector.

Main background in the vertex detector are \EPEM\ pairs produced in
\EPEM, \GE\ and \GG\ interactions. It is very similar to that in
\EPEM\ collisions and may differ only in numbers.
The shape of the zone occupied by the electrons kicked by the opposing beam is 
described by the formula  ~\cite{BATTEL},\cite{Brinkmann98}
\begin{equation}
r^2_{max} \simeq \frac{25Ne}{\sigma_zB}z \sim 0.12\frac{N}{10^{10}}
\frac{z\textrm{[cm]}}{\sigma_z\textrm{[mm]}B\textrm{[T]}},\;[\mbox{cm}^2]
\end{equation} 
where $r$ is the radius of the envelope at a distance 
$z$ from the IP, $B$ is longitudinal detector field.  For example, for
TESLA ($N=2\times10^{10}, \sigma_z=0.3$~mm, $B=3$~T) $r\textrm{[cm]} =
0.52\sqrt{z\textrm{[cm]}}$. 

So, from the background consideration it follows than the vertex
detector with a radius 2 cm can have the length about $\pm$ 15 cm and
a clear angle for the laser beams is $\pm$130 mrad.

As it was shown (see eq.\ref{thm}) the r.m.s. divergence of the the laser beam
at the conversion point $\sigma_{x^{\prime}}= 0.0155$; 2.5$\sigma$
will be sufficient. As we consider the optics situated outside the
electron beams, the required clear angle is $\pm 2\times 2.5 \times
0.0155 = \pm 78$ mrad.  As we see the laser beams have enough space inside
the vertex detector.

Together all the optics inside the detector covers the angle about
120--140 mrad (Fig.\ref{optics}). However, it does not mean that this
region is lost for the experiment. The thickness of mirrors will be
of the order of 1 $X_0$, and that will not affect too much the
performance of the calorimeter placed behind the mirrors.

\section{Conclusion}

Due to the potential decrease of the horizontal beam emittance in the
TESLA damping ring the luminosity in \GG\ collisions (in the high
energy peak) can reach about 1/3 of \EPEM\ luminosity. Since cross
sections in \GG\ collisions are typically higher by one order of
magnitude than those in \EPEM\ collisions and because \GG, \GE\ 
collisions give access to higher masses of some particles, the photon
collider now has very serious physics motivation.

The key problem for photon colliders is the powerful laser
system.  There are several possible laser-optics schemes for photon collider
at TESLA.  One of them has no visible problems and, it seems, can be
build now.

\section*{Acknowledgments}
I would like to thank K. Van Bibber, E.~Boos, R.~Brinkmann,
W.~Decking, A.~Gamp, M.~Galynskii, J.~Gronberg, A.~De Roeck,
I.F.~Ginzburg, K.~Hagiwara, R.~Heuer, C.~Heusch, V.~Ilyin, G.~Jikia,
 M.~Krawczyk, J.Kwiecinski, D.~Miller, M.~Perry, S.~Soldner-Rembold,
T.~Rizzo, W.~Sandner, S.~Schreiber, V.~Serbo, A.~Skrinsky,
T.~Takahashi, D.~Trines, A.~Wagner, I.~Will, N.~Walker, I.~Watanabe,
M.~Yurkov, K.~Yokoya, P.~Zerwas and all participants of the GG2000
workshop for useful discussions and support of photon colliders.


\begin{thebibliography}{99}
%
\bibitem{NLC} {\it Zeroth-Order Design Report for the Next Linear Collider} 
LBNL-PUB-5424, SLAC Report 474, May 1996.

\bibitem{JLC} {\it JLC Design Study}, KEK-REP-97-1, April 1997;
               I. Watanabe et. al.,KEK Report 97-17.


\bibitem{TESLA} {\it Conceptual Design of a 500 GeV Electron Positron
  Linear Collider with Integrated X-Ray Laser Facility} DESY 97-048,
  ECFA-97-182.
%  
\bibitem{CLIC}   R.W.~Assmann {\it et al.},
    ``A 3-TeV \EPEM\  linear collider based on CLIC technology,''
    CERN-2000-008.

\bibitem{GKST81} I.~Ginzburg, G.~Kotkin, V.~Serbo, V.~Telnov, {\it
Pizma ZhETF}, {\bf 34} (1981) 514; {\it JETP Lett.} {\bf 34} (1982)
491.


\bibitem{GKST83} I.~Ginzburg, G.~Kotkin, V.~Serbo, V.~Telnov, {\it Nucl. Instr.
and Meth.} {\bf 205} (1983) 47 (Prepr. INP 81-102, Novosibirsk, 1991).

\bibitem{GKST84} I.~Ginzburg, G.~Kotkin, S.~Panfil, V.~Serbo, V.~Telnov,
{\it Nucl. Instr. and Meth.} {\bf 219} (1984) 5 (Prepr. INP
 82-160, Novosibirsk, 1982).


\bibitem{TEL90} V.~Telnov, {\it Nucl. Instr. and Meth.} {\bf A 294}
  (1990) 72.

\bibitem{TEL95} V.~Telnov,
 {\it Nucl. Instr. Meth.} {\bf A 355} (1995) 3.


\bibitem{BERK} {\it Proc. of Workshop on \GG\ Colliders},
Berkeley CA, USA, 1994, {\it Nucl. Instr. {\rm\&}Meth.}
 {\bf A 355} (1995).

\bibitem{Brinkmann98} R.~Brinkmann et al., {\it Nucl. Instr. and Meth. }
  {\bf A 406} (1998) 13.

\bibitem{TELgoals} V.~Telnov, Goals of the Workshop GG2000, talk at
 {\it Intern. Workshop on High Energy Photon Colliders}, Hamburg, Germany,
 June 14-17, 2000, to be published in Nucl.~Instr.~Meth.~A (these proceedings).

\bibitem{Takahashi} T.~Takahashi, talk at {\it Intern. Workshop on High
 Energy Photon Colliders}, Hamburg, Germany, June 14-17, 2000, to be
 published in Nucl.~Instr.~Meth.~A (these proceedings).

  
\bibitem{Grutu} A.~Grutu, talk at {\it ICHEP2000}, Osaka, Japan, 27 July --
  2 August, 2000.

\bibitem{Okun} L.B.~Okun, Leptons and Quarks, North-Holand, Amsterdam, 1982.

\bibitem{ee97} V.~Telnov, {\it Int. J. Mod. Phys.} {\bf A 13} (1998) 2399,
hep-ex/9802003.

\bibitem{Bataglia} M.~Battaglia, HU-P-264, Apr 1999, {\it Proc. of 4th
  Intern. Workshop on Physics and Experiments at Linear Colliders
  (LCWS99)}, Sitges, Barcelona, Spain, 28 Apr - 5 May 1999,
  hep-ph/9910271.

\bibitem{Rembold99} G.~Jikia, S.~Soldner-Rembold, {\it Proc. of 4th
  International Workshop on Physics and Experiments at Linear
  Colliders (LCWS 99)}, Sitges, Barcelona, Spain, 28 Apr - 5 May 1999.
  e-print: hep-ph/9910366.

\bibitem{Melles} M.~Melles, W.J.~Stirling, V.A.~Khoze, 
{\it Phys. Rev.} {\bf D61} (2000) 054015,   e-print: hep-ph/9907238.

\bibitem{Rembold00} S.~Soldner-Rembold, G.~Jikia, talk at
 {\it Intern. Workshop on High Energy Photon Colliders}, Hamburg, Germany,
 June 14-17, 2000, to be published in Nucl.~Instr.~Meth.~A (these proceedings).

\bibitem{GinzKrav} I.~Ginzburg, M.~Krawczyk, {\it Proc. of 4th
  International Workshop on Physics and Experiments at Linear Colliders
  (LCWS 99)}, Sitges, Barcelona, Spain, 28 Apr - 5 May 1999,
  hep-ph/9909455 and talk at {\it Intern. Workshop on High Energy Photon
  Colliders}, Hamburg, Germany, June 14-17, 2000, to be published in
  {\it Nucl.~Instr.~Meth}.~A (these proceedings).

\bibitem{Accomando} E.~Accomando et al., {\it Phys. Rep.} {\bf 299} (1998) 1.

\bibitem{Muhlleitner} M.M.~Muhlleitner, talk at {\it Intern. Workshop
 on High Energy Photon Colliders}, Hamburg, Germany, June 14-17, 2000,
 to be published in {\it Nucl.~Instr.~Meth}.~A (these proceedings).

\bibitem{Tfrei} V.~Telnov, {\it Proc.  of the International Conference on the 
Structure and Interactions of the Photon (Photon 99)}, Freiburg,
Germany, 23-27 May 1999,  {\it Nucl. Phys. Proc. Suppl.} {\bf B82} (2000) 359, 
e-print: hep-ex/9908005. 

\bibitem{Ginz2000} I.F.~Ginzburg, talk at {\it Intern. Workshop on High
 Energy Photon Colliders,} Hamburg, Germany, June 14-17, 2000, to be
 published in {\it Nucl.~Instr.~Meth.}~A (these proceedings).

\bibitem{Arkani} N.~Arkani-Hamed, S.~Dimopoulos, G.~Dvali. SLAC-PUB-7769,
 March 1998, {\it Phys. Lett.} {\bf B 429} (1998) 263,  hep-ph/9803315. 
%
\bibitem{RIZZO} T.~Rizzo, {\it Proceedings of 4th International
  Workshop on Physics and Experiments at Linear Colliders (LCWS 99)},
  Sitges, Barcelona, Spain, 28 Apr - 5 May 1999, SLAC-PUB-8204,
  e-Print: hep-ph/9907401; talk at {\it Intern. Workshop on High
 Energy Photon Colliders,} Hamburg, Germany, June 14-17, 2000, to be
 published in {\it Nucl.~Instr.~Meth.}~A (these proceedings), hep-ph/0008037. 

\bibitem{Ferrario} M.~Ferrario, talk at {\it Intern. Workshop on High
 Energy Photon Colliders,} Hamburg, Germany, June 14-17, 2000, to be
 published in {\it Nucl.~Instr.~Meth.}~A (these proceedings).

\bibitem{Decking} W.~Decking, talk at {\it Intern. Workshop on High
 Energy Photon Colliders,} Hamburg, Germany, June 14-17, 2000, to be
 published in {\it Nucl.~Instr.~Meth.~A} (these proceedings).
\bibitem{Oide} K.~Hirata, K.~Oide, B.~Zotter, {\it Phys. Lett.} {\bf B224}
 (1989) 437. 

\bibitem{Walker} N.~Walker, talk at {\it Intern. Workshop on High
 Energy Photon Colliders}, Hamburg, Germany, June 14-17, 2000, to be
 published in {\it Nucl.~Instr. and Meth.}~A (these proceedings).

\bibitem{Sery}   P.~Raimondi, A.~Seryi, SLAC-PUB-8460, May 2000.  
Submitted to {\it Phys. Rev. Lett.}.

\bibitem{Sery1} A.~Seryi, private communication.

\bibitem{Brinkmann99} R.~Brinkmann, TESLA 99-15, Sep. 1999, {\it
  Proceedings of 4th International Workshop on Physics and Experiments
  at Linear Colliders (LCWS 99)}, Sitges, Barcelona, Spain, 28 Apr--5
  May 1999.

\bibitem{Galynskii} M.~Galynskii, E.~Kuraev, M.~Levchuk, V.~Telnov,
  talk at {\it Intern. Workshop on High Energy Photon Colliders,} Hamburg,
  Germany, June 14-17, 2000, to be published in {\it Nucl.~Instr.~Meth.}~A
  (these proceedings), hep-ph/0012338.
 
\bibitem{Gunion} J.F.~Gunion and J.G.~Kelly, {\it Phys. Lett.} {\bf B333} (1994) 110.
  
\bibitem{Kramer} M.~Kramer, J.~Kuhn, M.~Strong and P.~Zerwas, {\it Z.Phys.}
  {\bf C64} (1994) 21. 

\bibitem{CHTEL} P.~Chen, V.~Telnov, {\it Phys. Rev. Lett.}, {\bf 63}
(1989) 1796.
%


\bibitem{TSB1} V.~Telnov, SLAC-PUB-7337, {\it Phys. Rev. Lett.}, {\bf 78}
  (1997) 4757, erratum ibid 80 (1998) 2747, e-print: hep-ex/9610008.

\bibitem{Monter} V.~Telnov, {\it Proc. Advanced ICFA Workshop on
Quantum aspects of beam physics,} Monterey, USA, 4-9 Jan. 1998, World
Scientific, p.173, e-print: hep-ex/9805002.

\bibitem{Tlasv1} V.~Telnov, {\it Proc. of Intern. Symp. on New Visions in
Laser-Beam Interactions,} October 11-15, 1999, Tokyo, Metropolitan
University Tokyo, Japan, {\it Nucl. Instr. and Meth.} {\bf A450} (2000) 63,
hep-ex/0001029.

\bibitem{Will} I.~Will, T.~Quast, H.~Redlin and W.~Sandner, talk at
  {\it Intern. Workshop on High Energy Photon Colliders}, Hamburg, Germany,
  June 14-17, 2000, to be published in {\it Nucl.~Instr.~Meth.}~A (these
  proceedings).


\bibitem{DFEL} J.~Andruszkow et al., DESY-00-066, May
  2000, e-print: physics/0006010.

\bibitem{Yurkov} E.L.~Saldin, E.A.~Schneidmiller and M.V.~Yurkov, talk
  at {\it Intern. Workshop on High Energy Photon Colliders}, Hamburg,
  Germany, June 14-17, 2000, to be published in {\it Nucl.~Instr.~Meth.}~A
  (these proceedings).


\bibitem{Serbo} V.G.~Serbo, talk at {\it Intern. Workshop on High Energy
 Photon Colliders,} Hamburg, Germany, June 14-17, 2000, to be published
 in {\it Nucl.~Instr.~Meth.}~A (these proceedings).

\bibitem{Berestetskii} V.B.~Berestetskii, E.M.~Lifshitz and L.P.~Pitaevskii,
{\em Quantum electrodynamics \/} (Pergamon Press, Oxford, 1982).

\bibitem{koechner} W.~Koechner, {\it Solid State Laser Engineering},
Springer-Verlag.

\bibitem{Gronberg} J.~Gronberg, talk at {\it Intern. Workshop on High
 Energy Photon Colliders,} Hamburg, Germany, June 14-17, 2000, to be
 published in {\it Nucl.~Instr.~Meth.}~A (these proceedings).


\bibitem{Payne} S.A.~Payne, C.~Bibeau, C.D.~Marshall, H.T. Powell,
  UCRL-JC-119366, preprint LLNL, Dec.1998.



\bibitem{BATTEL}  M.~Battaglia, A.~Andreazza, M.~Caccia,
  V.Telnov, HIP-1997-522-exp, 1997; {\it Proc. of 2nd Workshop on
  Backgrounds at Machine Detector Interface, Honolulu}, HI, 21--22 Mar
  1997.  




\end{thebibliography}
\end{document}